\documentclass[eprintnumbers,amsmath,amssymb,twocolumn, showpacs, prl]{revtex4}
\usepackage{amsfonts}
\usepackage{mathrsfs}
\usepackage{graphicx}% Include figure files
\usepackage{dcolumn}% Align table columns on decimal point
\usepackage{bm}% bold math
\usepackage{times,epsfig,amssymb,amsmath}

\begin{document}

\title{No-compressing of quantum phase information}

\author{Yi-Nan Wang$^1$, Han-Duo Shi$^1$, Li Jing$^1$, Zhao-Xi Xiong$^1$, Jin Lei$^1$,
Liang-Zhu Mu$^1$\footnote{muliangzhu@pku.edu.cn}, and Heng
Fan$^2$\footnote{hfan@iphy.ac.cn}}
\affiliation{%
$^1$School of Physics, Peking University, Beijing 100871, China\\
$^2$Institute of Physics, Chinese Academy of Sciences, Beijing
100190, China
}%
\date{\today}% It is always \today, today,

\begin{abstract}
We raise a general question of quantum information theory whether
the quantum phase information can be compressed and retrieved. A
general qubit contains both amplitude and phase information, while
an equatorial qubit contains only a phase information. We study
whether it is possible to compress the phase information of $n$
equatorial qubits into $m$ general qubits with $m$ being less than
$n$, and still those information can be retrieved perfectly. We
prove that this process is not allowed by quantum mechanics.
\end{abstract}

\date{\today}

\maketitle

In quantum information processing, we can perform a lot of
miraculous tasks by using the principles of quantum mechanics, such
as, we can teleport an unknown quantum state \cite{teleportation} by
using Einstein-Podolsky-Rosen (EPR) pair \cite{EPR}, we can have a
quantum computer which surpasses its classical counterpart
\cite{shoralg}, we can construct protocols of quantum key
distribution with unconditional security \cite{BB84,ekert}, etc. On
the other hand, some tasks are not allowed by quantum mechanics, for
example, quantum information can not be cloned perfectly
\cite{nocloning}, it is even not allowed by quantum mechanics to
delete an unknown quantum state \cite{nodeleting}, it is impossible
to design a secure protocol of quantum bit commitment
\cite{bitcommitment0,bitcommitment}, the superluminal communication
is forbidden \cite{nosignaling}, etc.

It is continuously of broad interest and fundamental to explore the
realm of quantum mechanics to find what is possible and what is
impossible. Let us start with a problem of classical case: a
sequence of 10 bit can encode $2^{10}=1024$ different information,
while if the number of zeros and ones are equal in this bit
sequence, it can only encode $10!/(5!5!)=252$ different information.
Then the information represented in this special 10 bit sequence can
be compressed into a general $8$ bit sequence, since $2^8=256>252$.
Next let us consider the case of quantum information. A qubit
contains both amplitude and phase information which can be
represented explicitly in a Bloch sphere by two angles $\theta $ and
$\phi $,
\begin{eqnarray}
|\varphi \rangle =\cos \frac {\theta }{2}|0\rangle +\sin \frac {\theta }{2}e ^{i\phi }|1\rangle ,
\end{eqnarray}
where $\theta \in [0, \pi ], \phi \in [0,2\pi )$.
If we consider a specified qubit located in the equator of the Bloch sphere, we have an
equatorial qubit,
\begin{eqnarray}
|\psi \rangle = \frac {1}{\sqrt{2}}\left( |0\rangle +e ^{i\phi }|1\rangle \right),
\end{eqnarray}
there is only one phase parameter $\phi $. Of course, we know that by a
unitary transformation similar to the Hadamard gate, $U_r$ which will be
presented later in (\ref{ur}) and its inverse $U_r^{\dagger }$, the
phase parameter and the amplitude parameter can be exchanged. A
general qubit and an equatorial qubit are different. For example, in
universal quantum cloning, the optimal fidelity for qubit can be
around $83.3\% $, \cite{UQCM}, while for equatorial qubit, the
optimal fidelity can be higher and achieves about $85.4\% $,
\cite{phase-clone,fanphase}.

We can have another example: suppose we would like to teleport an
unknown two-qubit state but with a partially known form, $|\Psi
\rangle =\alpha |00\rangle +\beta |11\rangle $. Instead of using two
EPR pairs, we can {\it extract} the information in this two-qubit
state by a local controlled-NOT (CNOT) gate applied on $|\Psi
\rangle$ with the first qubit as the controlled qubit and the second
as the target gate. Alice, the sender, then obtains, $|\Psi_A
\rangle=(\alpha |0 \rangle + \beta | 1 \rangle) | 0 \rangle$. Alice
only needs to teleport the first qubit by a resource of one EPR pair
\cite{teleportation}, Bob, the receiver, would receive $|\psi
\rangle = \alpha | 0 \rangle + \beta | 1 \rangle$ by teleportation.
By adding an ancillary state $|0\rangle$, Bob can locally perform a
CNOT gate and finally recover the original two-qubit state, $|\Psi
\rangle =\alpha |00\rangle +\beta |11\rangle $.  So we are able to
{\it compress} the information of a partially known two-qubit state
to a single qubit, such that we can teleport it by a  resource of
only one EPR pair.

The fact that we can teleport a two-qubit state, $|\Psi \rangle
=\alpha|00\rangle + \beta |11\rangle$, by only one EPR pair is due
to that this two-qubit state, or we can rewrite it as, $|\Psi
\rangle =\cos \frac{\theta}{2} | 00 \rangle + \sin \frac{\theta}{2}
e^{i \phi} | 11 \rangle$, contains only the information of two
angles, namely $\theta$ and $\phi$, just as an ordinary qubit $|
\varphi \rangle = \cos \frac{\theta}{2} | 0 \rangle + \sin
\frac{\theta}{2} e^{i \phi} | 1 \rangle$ does. We may then ask a
question in an opposite direction: since each equatorial qubit has a
fixed $\theta =\pi /2$, and contains only the information of a
single angle, $\phi$, is it possible to {\it compress} the
information of two angles of two equatorial qubits into just one
general qubit? If yes, we can teleport it by only one EPR pair, then
can we {\it separate} the information from this qubit into two
equatorial qubits and recover their original form (We assume that only
reversible operations are used in the recovery process,
hence measurement is not allowed)? We will next
prove that this procedure is not allowed by quantum mechanics!

Let us start with a simple example. Consider two equatorial qubits as the following,
\begin{eqnarray}
|\psi_1 \rangle = \frac{1}{\sqrt{2}} ( | 0 \rangle + e^{i \phi_1} |1 \rangle) \\
|\psi_2 \rangle = \frac{1}{\sqrt{2}} ( | 0 \rangle + e^{i \phi_2} |1 \rangle)
\end{eqnarray}
Naturally, we might rotate one of the equatorial qubits so that the
equatorial large circle would be rotated to a longitudinal one by unitary
transformation,
\begin{eqnarray}
U_r=\frac{1}{\sqrt{2}}
\left(
\begin{array}{cc}
1 & -1 \\
-i & -i
\end{array}
\right) , \label{ur}
\end{eqnarray}
so we have,
\begin{eqnarray}
U_r | \psi_1 \rangle = \sin \frac{\phi_1}{2} | 0 \rangle + \cos \frac{\phi_1}{2} | 1 \rangle .
\end{eqnarray}
Now, the angle of phase information is changed to the angle of amplitude information.
For the two qubits, $(U_r| \psi_1 \rangle )| \psi_2 \rangle$,
apply a CNOT gate to the system with the first
qubit as the controlled qubit and the second as the target gate, and
we would obtain,
\begin{eqnarray}
| \Phi \rangle &=& \frac{1}{\sqrt{2}} [ (\cos \frac{\phi_1}{2} | 1 \rangle
+ \sin \frac{\phi_1}{2} e^{i \phi_2} | 0 \rangle) | 1 \rangle
\nonumber \\
&&+ (\sin \frac{\phi_1}{2} | 0 \rangle + \cos \frac{\phi_1}{2} e^{i
\phi_2} | 1 \rangle) | 0 \rangle].
\end{eqnarray}
It is now clear,
by measuring the second qubit, for both $|0\rangle $ and $|1\rangle
$ cases, we would get a single qubit containing the information of
both $\phi_1$ and $\phi_2$,
\begin{eqnarray}
| \varphi _1 \rangle = \cos \frac{\phi_1}{2}
| 1 \rangle + \sin \frac{\phi_1}{2} e^{i \phi_2} | 0 \rangle,\\
| \varphi _2 \rangle = \sin \frac{\phi_1}{2} | 0 \rangle +
\cos \frac{\phi_1}{2} e^{i\phi_2} | 1 \rangle.
\end{eqnarray}
It is now possible to teleport one qubit with two angles information
by one EPR pair. Then it is Bob's problem:  is it possible to
separate the phase and amplitude information from  $| \varphi
_{1(2)} \rangle$? However, for example in case, $| \varphi
_1\rangle$, Bob would find that if $\phi_1 = 0,\pi $, the phase
information in $|\varphi_1\rangle $ would be totally lost. Then we
conclude that it is impossible to recover the original two
equatorial qubits. This procedure is represented in Fig. 1. We shall
provide a proof later for the general case.

We may also understand this in the language of topology. The initial
state constructs a space that is the direct product of two
one-dimensional rings $S^1 \times S^1$. And this space is
homeomorphic to the ordinary torus surface in our three-dimensional
space. However, the state, $|\varphi _1\rangle =\cos
\frac{\phi_1}{\sqrt{2}}
 | 1 \rangle + \sin \frac{\phi_1}{2} e^{i \phi_2} | 0 \rangle$,
forms a two-dimensional sphere $S^2$.  Since these two
spaces are not homeomorphic, there exists
no bijective continuous mapping from one space to another.

\begin{figure}
\includegraphics[height=6.15cm,width=7cm]{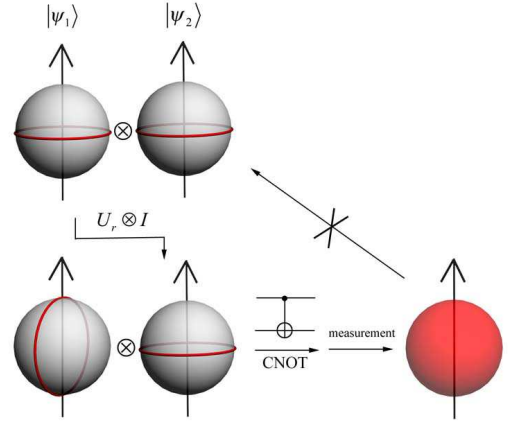}
\caption{(color online)An equatorial qubit is located in the equator of
the Bloch sphere. Two equatorial qubits by one single qubit rotation $U_r$ and
a CNOT gate followed by a measurement can lead to a qubit with two angles information.
However, it is impossible to recover the original form of two equatorial qubits
from just one single qubit with both phase and amplitude information. }
\end{figure}

Now let us consider a general case. One may find that the problem of compressing and
retrieving of quantum phase information is whether it is possible by unitary transformation
to retrieve the information of $m$ qubits into $n$ equatorial qubits, $n>m$.
Or similarly, whether a unitary transformation can change $n$ equatorial qubits
to $m$ qubits,
\begin{eqnarray}
U\bigotimes _{k=1}^n|\psi _k\rangle |A\rangle =\bigotimes _{k=1}^m|\varphi _k\rangle |B\rangle ,
\end{eqnarray}
where, $|\psi _k\rangle =\frac{1}{\sqrt{2}} ( | 0 \rangle + e^{i \phi_k} |1 \rangle)$, is the equatorial
qubit, and  $|\varphi _k\rangle $ is a general qubit, $|A\rangle ,|B\rangle $ are ancillary states.
We can let $\phi _k=0,\pi $, so we have $|\psi _k\rangle =(|0\rangle \pm |1\rangle )/\sqrt {2}$, there are
altogether $2^n$ orthogonal states, while $m$ general qubits can provide $2^m$ orthogonal states.
We know that unitary transformations keep orthogonal states to be orthogonal, when $n>m$, the
unitary operator $U$ does not exist. Thus the procedure of compressing and retrieving
quantum phase information is impossible.

For a more general d-dimension case, we also have a similar result.
Define the d-dimension equatorial state as,
$\sum_{j=0}^{d-1}e^{i\phi_{k_j}}|j\rangle $, where the normalization
factor $\frac{1}{\sqrt{d}}$ is omitted hereafter. It forms a
(d-1)-dimensional torus $T^{d-1}=S^1\times\cdots \times S^1$, the
direct product of (d-1) one-dimensional rings. The $n$ input states
can be expressed as,
\begin{eqnarray}
\psi_{in}=\bigotimes_{k=1}^{n}(\sum_{j=0}^{d-1}e^{i\phi_{k_j}}|j\rangle).
\end{eqnarray}
The question is, whether there exists a unitary transformation, $U$, satisfying
the following equation:
\begin{eqnarray}
U(\psi_{in}\otimes\psi_A)=\psi(\phi_{1_0},\cdots,\phi_{n_{d-1}})\otimes\psi_B.
\label{unitary}
\end{eqnarray}
where $\psi_A\in \mathcal{H}^{d^p}$ and $\psi_B\in
\mathcal{H}^{d^{n-m+p}}$ denotes ancillary states independent of the
input parameters, and $\psi(\phi_{1_0},\dots,\phi_{n_{d-1}})\in
\mathcal{H}^{d^m}$ is the compressed state. If such a unitary matrix
exists, we can invert the whole process to finish the retrieving
process and recover the original state. Without loss of generality,
we suppose $\psi_A=|0\rangle^p$ and $\psi_B=|0\rangle^{(n-m+p)}$.
Then, the equation (\ref{unitary}) can be rewritten as,
\begin{eqnarray}
&&U[\bigotimes_{k=1}^{n}(\sum_{j=0}^{d-1}e^{i\phi_{k_j}}|j\rangle)\otimes|0\rangle^{\otimes
p}]
\nonumber \\
&=& \psi(\phi_{1_0},\cdots,\phi_{n_{d-1}})\otimes|0\rangle^{\otimes
(p+n-m)}.
\end{eqnarray}
Clearly, the final superposition state contains no term in the form,
$|a_{1}a_{2}\cdots a_{n+p}\rangle$, with
$a_{m+1}$,$a_{m+2}$,$\cdots$,$a_{n+p}$ are not all zeros. So the
coefficients of the unitary matrix $U$, $u_{i_1i_2\cdots
i_{n+p},j_1j_2\cdots j_{n+p}}$, must obey the following relation,
\begin{eqnarray}
\sum_{j_1=0}^{d-1}\sum_{j_2=0}^{d-1}\cdots\sum_{j_n=0}^{d-1}\prod_{k=1}^{n}e^{i\phi_{k_j}}u_{a_1\cdots
a_{n+p},j_1j_2\cdots j_{n}00\cdots 0}=0, \label{u-coefficient}
\end{eqnarray}
where, as we mentioned, $a_{m+1}$,$a_{m+2}$,$\cdots$,$a_{n+p}$ are
not all zeros.

Lemma 1: {\it If for any arbitrary
$\phi_{1_0}\dots\phi_{n_{d-1}}\in[0,2\pi]$,
\begin{eqnarray}
\sum_{j_1=0}^{d-1}\sum_{j_2=0}^{d-1}\cdots\sum_{j_n=0}^{d-1}\prod_{k=1}^{n}e^{i\phi_{k_j}}x_{j_1j_2\cdots
j_n}=0,
\end{eqnarray}
then each of the coefficient, $x_{j_1\dots j_n}=0$.}\\
\noindent \emph {Proof}: We set each parameter $\phi_{k_1}$ to one of the $d$ possible values,
$0,\frac{2\pi}{d},\frac{4\pi}{d},\cdots,\frac{2(d-1)\pi}{d}$, and we
set $\phi_{k_j}=j\phi_{k_1}$ for every $0\le j\le d-1$ and
every $1\le k\le n$, then $d^n$ equations are formed.
%Namely,
%in the $y$-th equation, if $(y-1)$ can be written as a d-ary number
%$y_1\dots y_n$, we set $\phi_{k_1}=\frac{2s\pi}{d}$ if $y_k=s$, for
%every $1\le k\le n$ and every $1\le y\le d^n$.

Now we denotes the coefficient matrix of these $d^n$ equations as
$A_n$. We only have to prove that $|A_n|\neq 0$. According to the
above construction method, we have
\begin{eqnarray}A_1=
\begin{pmatrix}
1  &  1  &  1  &  \cdots  &  1  \\
1  &  \omega  &  \omega^2  &  \cdots  &  \omega^{d-1}  \\
1  &  \omega^2  &  \omega^4  &  \cdots  &  \omega^{2(d-1)}  \\
\vdots  &  \vdots  &  \vdots  &  \ddots  &  \vdots  \\
1  &  \omega^{d-1}  &  \omega^{2(d-1)}  &  \cdots  &  \omega^{(d-1)^2}  \\
\end{pmatrix},
\end{eqnarray}
where $\omega=e^{\frac{2\pi i}{d}}$. It is a Vandermonde matrix, so
its determinant takes the form,
\begin{eqnarray}
|A_1|=\prod_{0\leqslant k  <  l\leqslant (d-1)}(\omega^k-\omega^l).
\end{eqnarray}
And we also have,
\begin{eqnarray}
A_{n+1}=
\begin{pmatrix}
A_n  &  A_n  &  A_n &  \cdots  &  A_n  \\
A_n  &  A_n\omega  &  A_n\omega^2  &  \cdots  &  A_n\omega^{d-1}  \\
A_n  &  A_n\omega^2  &  A_n\omega^4  &  \cdots  &  A_n\omega^{2(d-1)}  \\
\vdots  &  \vdots  &  \vdots  &  \ddots  &  \vdots  \\
A_n  &  A_n\omega^{d-1}  &  A_n\omega^{2(d-1)}  &  \cdots  &  A_n\omega^{(d-1)^2}  \\
\end{pmatrix},
\end{eqnarray}
\begin{eqnarray}
|A_N|=[\prod_{0\leqslant k  <  l\leqslant (d-1)}(\omega^k-\omega^l)]^N.
\end{eqnarray}
While $k<l$, $\omega^k\neq \omega^l$, hence these determinants are
not zero. Q.E.D.

%We use ``0-row'' to denote a row in the unitary matrix $U$ if the
%d-ary row-number $j_1\dots j_{n+p}$ ends up with $p$
%zeros.($j_{n+1}=\dots =j_{n+p}=0$). Obviously, there're $d^n$ such
%kind of rows.

Coming back to the matrix $U$ in (\ref{u-coefficient}), from Lemma
1, each column consists at most $d^m$ non-zero terms, and these
non-zero terms are on the same rows. We may interchange the rows and
swap the columns to have the matrix $U$ like the following form,

\begin{eqnarray}
U=
\begin{pmatrix}
u_{a_{11}}  &  u_{a_{12}}  &  \cdots  & u_{a_{1d^n}}  &  \cdots   & u_{a_{1d^{n+p}}}  \\
\vdots  &  \vdots  &  \ddots  &  \vdots   &   \ddots  & \vdots        \\
u_{a_{d^{m}1}}  &  u_{a_{d^{m}2}}  &  \cdots  & u_{a_{d^{m}d^n}}  &  \cdots   & u_{a_{d^{m}d^{n+p}}}  \\
0 &  0  &  \ldots &  0  &\cdots&   \\
\vdots  &  \vdots  &  \ddots  &  \vdots  &  &  \\
0 &  0  &  \ldots &  0  &\cdots&   \\
\end{pmatrix}
\end{eqnarray}
It is a plain fact that if $m<n$, $|U|=0$. This contradicts the
proposition that $U$ is a unitary matrix. We thus conclude that the
compressing and retrieving of the quantum phase information is not
allowed by quantum mechanics.

\emph{Discussions.}---As for no-cloning theorem \cite{nocloning}, we
may not clone an un-known quantum state perfectly, however, we can
try to clone it approximately \cite{UQCM,N_UQCM,Werner,Fan} or
probabilistically \cite{DG}. While we find that the quantum phase
information can not be compressed and retrieved perfectly, it may
still be interesting to design protocols for different purposes so
that the quantum phase information can be compressed in some other
senses.

We also would like to point out that the quantum information
compressing has already been widely studied in quantum coding
theories, see for example Refs.\cite{schum,q-compress}, where the
ensemble of signals from the signal source represented by density
matrix is considered. However our compressing of the quantum phase
information in this Letter is in a completely different framework.

A global phase of a quantum state in general cannot be detected, only the
relative phase has the physical meaning, it needs at least two energy levels
to encode it. This might be the simplest
no-compressing of the quantum phase information. So the problem
studied in this paper should be fundamental for any physical systems.

In summary, supplementary to those well-known impossibilities for
quantum information theory
\cite{nocloning,nodeleting,bitcommitment0,bitcommitment,nosignaling},
we show that the compressing and retrieving of quantum phase information
is not allowed by quantum mechanics. Similar as for other
impossibility cases, it is also expected that we may design some
other phase information compressing protocols and it would also be
interesting to relate our result with some other fundamental
theorems in quantum information and quantum mechanics.

This work is supported by NSFC (10974247, 11175248), ``973'' program (2010CB922904), NFFTBS (J1030310).

\end{document}